\documentclass[preprint,aps]{revtex4}
\usepackage{amsmath}
\usepackage{amssymb}

\input{tcilatex}

\begin{document}

\title{The analytical description of a doped Mott insulator}
\author{Yu-Liang Liu}
\affiliation{Department of Physics, Renmin University, Beijing 100872, \\
People's Republic of China}

\begin{abstract}
With the hierarchical Green's function approach, we study a doped Mott
insulator described with the Hubbard model by analytically solving the
equations of motion of an one-particle Green's function and related
multiple-point correlation functions, and find that the separation of the
spin and charge degrees of freedom of the electrons is an intrinsic
character of the doped Mott insulator. For enough of large on-site repulsive
Coulomb interaction, we show that the spectral weight of the one-particle
Green's function is proportional to the hole doping concentration that is
mainly produced by the charge fluctuation of electrons, while the excitation
spectrum of the electrons is composed of two parts: one is contributed by
the spin fluctuation of the electrons which is proportional to the hole
doping concentration, and another one is coming from the coupling between
the charge and spin fluctuations of the electrons that takes the maximum at
undoping. All of these low energy/temperature physical properties originate
from the strong on-site Coulomb interaction. The present results are
consistent with the spectroscopy observations of the cuprate
superconductors, and the numerical calculations in normal state above
pseudogap regime.
\end{abstract}

\maketitle

\section{\protect\bigskip Introduction}

Since the discovery of the high $T_{c}$ cuprate superconductive materials%
\cite{1}, it is gradually realized that the strong correlation effect of
electrons play a key role in understanding of the normal and superconducting
states of these materials\cite{2,3,4,5}. Up to now there are a lot of
experimental data and numerical simulations showing that the novel behavior
of the normal states in the underdoped and optimal doped regimes of these
materials\cite{5a,6,7} originates from the strong correlation of electrons
produced by the strong on-site repulsive Coulomb interaction of electrons,
and these unprecedented properties cannot be unambiguously explained by
usual perturbation theory of quantum many particle systems based on the
"independent particle" (quasi-particle) assumption of the Landau Fermi
liquid theory\cite{8}.

The Hubbard model and the related t-J model are widely thought to capture
the essential physics of a class of highly correlated systems, such as the
high T$_{c}$ cuprate superconductors. The two-dimensional (2D) Hubbard model%
\cite{7a} on a square lattice, used to describe the basic characters of high 
$T_{c}$ cuprate supercunductivity\cite{5a,6,7,8a,88a}, has been extensively
studied in both analytical and numerical calculations, where there is
inherent frustration between the tendency to maintain local
antiferromagnetic correlations originated from the strong on-site repulsive
Coulomb interaction and the doped hole itineracy.

The effective treatment of the influence of the on-site repulsive Coulomb
interaction on the states of electrons is a central issue of any theoretical
approach, where at the large repulsive $U$, a double occupied state on each
site is strongly suppressed, and the Hilbert space of the electrons is split
into two subspaces: one is composed of the unoccupied and single occupied
states, and another one composed of the double occupied states that are
lifted up high energy levels. In fact, there emerges a single-occupied
constraint condition for electrons on each site produced by the strong
on-site repulsive Coulomb interaction, which is a major difficulty faced by
the present approaches. On the other hand, it is well known that in the both
cases of weak $U/t_{0}\ll 1$ and strong $U/t_{0}\rightarrow \infty $
coupling limits, where $t_{0}$ is the hopping amplitude of electrons, the
basic property of the ground state of the 2D square lattice Hubbard model is
clear: in the former it is a Fermi liquid\cite{81}, and in the latter it is
a fully polarized ferromagnetic metallic phase\cite{82} away from the half
filling, in which there does not appear any order state.

The rich physical phenomena shown by the 2D square lattice Hubbard model
really appear in the intermediate coupling, where $U$ is of order the
bandwidth $W$($=8t_{0}$), $U\sim W$, and there is the keen competition
between the kinetic energy and the on-site repulsive Coulomb interaction of
electrons. The former takes the delocalization of electrons, while the
latter makes electrons localize. In this coupling range, there is still not
a ubiquitous acceptable calculation from microscopic theories. The 2D square
lattice Hubbard model with intermediate coupling, likely cannot be treated
using any fundamentally perturbation approach which starts with a
non-interacting particle description because the strength of the
interactions among electrons is comparable to or large than their kinetic
energy. That is that there is not any controllable effectively small
"interaction strength" as a perturbation expansion parameter in this system
due to the strong correlation among electrons. Beyond the present
perturbation theoretical methods, the on-site Coulomb interaction of
electrons had to be treated effectively before taking any approximation in
analytical and numerical calculations.

The analytical description of a strongly correlated system is very
successful only for one-dimensional case, such as the Bethe ansatz\cite%
{9,99b,9a} and bosonization method\cite{9aa} that cannot be extended for two
or higher dimensions. However, any theory based on a perturbation expansion%
\cite{9b,9c} around the non-interacting limit is at least questionable, due
to the non-perturbation nature of the strongly correlated system\cite%
{99a,99c,99d,99e}. Beyond usual equation of motion of Green's function
approach\cite{10,11,12,13,13a,13ab}, we use the hierarchic Green's function
approach\cite{13aa} to study the intrinsic character of a doped Mott
insulator with the Hubbard model under the strong on-site repulsive Coulomb
interaction.

For a doped Mott insulator, the central issue is rigorous and/or effective
treatment of the competition between the doped hole itineracy and the strong
on-site repulsive Coulomb interaction to maintain local antiferromagnetic
correlation of spins, that produces some new orders in the low
energy/temperature region. This competition in fact is a many body effect of
electrons and it cannot be effectively described by a perturbation parameter
like that in usual weakly correlated systems, such as three-dimensional
electron gas in high electron density limit. In this work, we make a step to
this end. For simplicity, without introducing the pseudogap and other order
parameters, we mainly study the influence of the doped holes on the
one-particle Green's function and the low-lying excitation spectrum under
the large on-site Coulomb interaction $U$ which is of order the bandwidth $W$%
; Then we demonstrate that for the small hole doping concentration $\delta $%
, the spectral weight of the electrons is proportional to $\delta $, that is
mainly attributed to the charge fluctuation of electrons, while the
excitation spectrum is composed of two parts: one is mainly contributed by
the spin fluctuation of the electrons, which is proportional to $\delta $,
and another one originates from the coupling between the charge and spin
fluctuations of the electrons, which is proportional to $1-\delta $. In
fact, there takes place the separation of the charge and spin degrees of
freedom of electrons. In Sec.II, with the hierarchic Green's function
approach, we give some key equations of motion of one-particle Green's
function and related multiple-point correlation functions. In Sec. III,
under the soft cut-off approximation (see below), we solve these equations
of motion, and give an analytic expression of the one-particle Green's
function. With these solutions, we demonstrate that the spectral weight of
electrons is proportional to the hole doping concentration $\delta $, and
the double occupation function of electrons goes to zero in the both regions
around $\delta =0$ and $\delta =1$, respectively. We give our conclusion and
discussion in Sec. IV, and more detail calculations in the Appendix.

\section{The equation of motion of the one-particle Green's function}

With the hierarchic Green's function approach, we can write out the equation
of motion of the one-particle Green's function (choosing $\hbar =1$),%
\begin{equation}
\left[ i\partial _{t}+\mu \right] G_{iq\sigma }(t)=\delta (t)\delta
_{iq}+\sum_{m}\widehat{h}_{im}G_{mq\sigma }(t)+\frac{U}{2}F_{iiq}^{(n)}(t)-%
\frac{\sigma U}{2}F_{iiq}^{(s)}(t)  \label{3a}
\end{equation}%
where, $F_{ijq}^{(n)}(t)=-i<\mathit{T}\widehat{n}_{i}\left( t\right) 
\widehat{c}_{\sigma j}(t)\widehat{c}_{\sigma q}^{\dagger }(0)>$, and $%
F_{ijq}^{(s)}=-i<\mathit{T}\widehat{s}_{i}\left( t\right) \widehat{c}%
_{\sigma j}(t)\widehat{c}_{\sigma q}^{\dagger }(0)>$, they are produced by
the on-site Coulomb interaction of electrons. Here $\widehat{n}_{i}=\widehat{%
n}_{\uparrow i}+\widehat{n}_{\downarrow i}$ is the charge density operator,
and $\widehat{s}_{i}=\widehat{n}_{\uparrow i}-\widehat{n}_{\downarrow i}$ is
the spin density operator. The related multiple-point correlation function $%
F_{ijq}^{(n)}(t)$ represents the contribution of the charge fluctuation of
electrons to the one-particle Green's function, while the related
multiple-point correlation function $F_{ijq}^{(s)}(t)$ represents the
contribution of the spin fluctuation of electrons to the one-particle
Green's function. This linear equation of motion of the one-particle Green's
function is rigorous, and the strong correlation effect of electrons is
completely represented by the on-site correlation functions $%
F_{iiq}^{(n)}(t) $, and $F_{iiq}^{(s)}(t)$, thus these two correlation
functions play a key role in solving the equation of motion of the
one-particle Green's function. Here, we do not consider the influence of the
pseudogap of electrons on the low energy states, and the present calculation
is valid for a higher energy scale than the pseudogap $\Delta $ in the
underdoping region.

The equations of motion of the multiple-point correlation function $%
F_{ijq}^{(n)}(t)$ and $F_{ijq}^{(s)}(t)$ can be written as that,%
\begin{eqnarray}
\left[ i\partial _{t}+\mu \right] F_{ilq}^{(n)}(t) &=&<\widehat{n}%
_{i}>\delta (t)\delta _{lq}+\frac{U}{2}F_{illq}^{(nn)}(t)-\frac{\sigma U}{2}%
F_{illq}^{(ns)}(t)  \notag \\
&&+\sum_{m}\left[ \widehat{h}_{lm}F_{imq}^{(n)}(t)+\widehat{h}%
_{im}F_{imlq}^{(P^{\left( -\right) })}(t)\right]  \label{3a1}
\end{eqnarray}%
\begin{eqnarray}
\left[ i\partial _{t}+\mu \right] F_{ilq}^{(s)}(t) &=&<\widehat{s}%
_{i}>\delta (t)\delta _{lq}+\frac{U}{2}F_{illq}^{(sn)}(t)-\frac{\sigma U}{2}%
F_{illq}^{(ss)}(t)  \notag \\
&&+\sum_{m}\left[ \widehat{h}_{lm}F_{imq}^{(s)}(t)+\widehat{h}%
_{im}F_{imlq}^{(Q^{\left( -\right) })}(t)\right]  \label{3a2}
\end{eqnarray}%
where the definition of the high order related multiple-point correlation
functions appearing in the above equations is given in the Appendix. With
the same procedures, we can write out the equations of motion of these
correlation functions in which there will appear more high order new
correlation functions, and this set of equations of motion is not closed. In
order to calculate the one-particle Green's function, we need to cut-off
this set of equations of motion at some level. According to the above
equations, it is clear that the correlation functions $F_{ilq}^{(n)}(t)$ and 
$F_{ilq}^{(s)}(t)$ are directly coupled by the high order correlation
function $F_{ijlq}^{(sn)}(t)$, while they are independent of each other as
without completely considering the contribution of these high order
correlation functions. In the large on-site repulsive Coulomb interaction
region, the coupling between both the charge and spin fluctuation plays
important role in studying of the low energy behavior of the one-particle
Green's function, and the following approximation taken would effectively
incorporate in the contribution of the correlation function $%
F_{ijlq}^{(sn)}(t)$. Here the correlation functions $F_{ijlq}^{(P^{\left(
-\right) })}(t)$ and $F_{ijlq}^{(Q^{\left( -\right) })}(t)$ represent high
order charge and spin fluctuations, and their contribution to the
correlation functions $F_{ilq}^{(n)}(t)$ and $F_{ilq}^{(s)}(t)$ is to mainly
modify the chemical potential.

The Eqs.(\ref{3a}-\ref{3a2}) are the key ones that we use them to calculate
the one-particle Green's function in the large $U$ region, where the spin
and charge fluctuations are separated obviously that are represented by the
multiple-point correlation functions $F_{ilq}^{(n)}(t)$ and $%
F_{ilq}^{(s)}(t) $, respectively. Next we write out the equations of motion
of the high order multiple-point correlation functions appearing in Eqs.(\ref%
{3a}-\ref{3a2}),%
\begin{eqnarray}
\left[ \omega +\mu \right] F_{iljq}^{(nn)}(\omega ) &=&<\widehat{n}_{i}%
\widehat{n}_{l}>\delta _{jq}+\frac{U}{2}F_{iljjq}^{(nnn)}(\omega )-\frac{%
\sigma U}{2}F_{iljjq}^{(nns)}(\omega )  \notag \\
&&+\sum_{m}\left[ \widehat{h}_{jm}F_{ilmq}^{(nn)}(\omega )+\widehat{h}%
_{im}F_{imljq}^{(P^{\left( -\right) }n)}(\omega )+\widehat{h}%
_{lm}F_{ilmjq}^{(nP^{\left( -\right) })}(\omega )\right]  \label{3a3}
\end{eqnarray}%
\begin{eqnarray}
\left[ \omega +\mu \right] F_{iljq}^{(sn)}(\omega ) &=&<\widehat{s}_{i}%
\widehat{n}_{l}>\delta _{jq}+\frac{U}{2}F_{iljjq}^{(snn)}(\omega )-\frac{%
\sigma U}{2}F_{iljjq}^{(sns)}(\omega )  \notag \\
&&+\sum_{m}\left[ \widehat{h}_{jm}F_{ilmq}^{(sn)}(\omega )+\widehat{h}%
_{im}F_{imljq}^{(Q^{\left( -\right) }n)}(\omega )+\widehat{h}%
_{lm}F_{ilmjq}^{(sP^{\left( -\right) })}(\omega )\right]  \label{3a4}
\end{eqnarray}%
\begin{eqnarray}
\left[ \omega +\mu \right] F_{iljq}^{(ss)}(\omega ) &=&<\widehat{s}_{i}%
\widehat{s}_{l}>\delta _{jq}+\frac{U}{2}F_{iljjq}^{(ssn)}(\omega )-\frac{%
\sigma U}{2}F_{iljjq}^{(sss)}(\omega )  \notag \\
&&+\sum_{m}\left[ \widehat{h}_{jm}F_{ilmq}^{(ss)}(\omega )+\widehat{h}%
_{im}F_{imljq}^{(Q^{\left( -\right) }s)}(\omega )+\widehat{h}%
_{lm}F_{ilmjq}^{(sQ^{\left( -\right) })}(\omega )\right]  \label{3a5}
\end{eqnarray}%
where there emerge new more higher order multiple-point correlation
functions, and there appear the static spin-spin and density-density
correlation functions that can be self-consistently determined by
calculating spin-spin and density-density correlation functions under
equal-time limit in the above equations. In order to effectively including
the coupling between the spin and charge fluctuation of electrons induced by
the large on-site repulsive Coulomb interaction, we only take some
approximations in the Eqs.(\ref{3a3}-\ref{3a5}). For simplicity, we neglect
all the static quantities appearing in these equations.

In the equation of motion of the one-particle Green's function Eq.(\ref{3a}%
), there only appears the correlation functions $F_{iiq}^{(n)}(t)$ and $%
F_{iiq}^{(s)}(t)$, while in their equations of motion Eq.(\ref{3a1},\ref{3a2}%
) (taking $l=i$) there emerge the correlation functions $%
F_{iiiq}^{(nn)}(t)=-i<\mathit{T}\left[ \widehat{n}_{i}\left( t\right) \right]
^{2}\widehat{c}_{\sigma i}(t)\widehat{c}_{\sigma q}^{\dagger }(0)>$ and $%
F_{iiiq}^{(ss)}(t)=-i<\mathit{T}\left[ \widehat{s}_{i}\left( t\right) \right]
^{2}\widehat{c}_{\sigma i}(t)\widehat{c}_{\sigma q}^{\dagger }(0)>$. On the
other hand, the correlation functions $F_{iiq}^{(n)}(t)/F_{iiq}^{(s)}(t)$
and $F_{ilq}^{(n)}(t)/F_{ilq}^{(s)}(t)$ are connected by the Eq.(\ref{3a1})/(%
\ref{3a2}), in which there emerge the correlation functions $%
F_{illq}^{(nn)}(t)$, $F_{illq}^{(sn)}(t)$, $F_{illq}^{(ss)}(t)$, and others.
As taking $l=j$ or $i=j$ in the Eqs.(\ref{3a3}-\ref{3a5}), there will be a
lot of the correlation functions that in their definitions there appear the
charge operator $\left[ \widehat{n}_{i}\left( t\right) \right] ^{2}$ or the
spin operator $\left[ \widehat{s}_{i}\left( t\right) \right] ^{2}$, for
example, $F_{ijjjq}^{(nnn)}(t)=-i<\mathit{T}\widehat{n}_{i}\left( t\right) %
\left[ \widehat{n}_{j}\left( t\right) \right] ^{2}\widehat{c}_{\sigma j}(t)%
\widehat{c}_{\sigma q}^{\dagger }(0)>$, and $F_{ijjjq}^{(sss)}(t)=-i<\mathit{%
T}\widehat{s}_{i}\left( t\right) \left[ \widehat{s}_{j}\left( t\right) %
\right] ^{2}\widehat{c}_{\sigma j}(t)\widehat{c}_{\sigma q}^{\dagger }(0)>$, 
\textit{et al}.. It is a key point how to effectively treat these
correlation functions that have the charge operator $\left[ \widehat{n}%
_{i}\left( t\right) \right] ^{2}$ or the spin operator $\left[ \widehat{s}%
_{i}\left( t\right) \right] ^{2}$ appearing in the Eqs.(\ref{3a1}-\ref{3a5}%
). In the following we use a special relation between the charge operator $%
\widehat{n}_{i}$ and spin operator $\widehat{s}_{i}$ to approximately treat
these correlation functions.

Instead of taking a simple cut-off approximation for high order
multiple-point correlation functions, we use the character of the charge
operator $\widehat{n}_{i}$ and spin operator $\widehat{s}_{i}$ to simplify
these equations, where they have the following relation,%
\begin{equation}
\left( \widehat{n}_{i}\right) ^{2}+\left( \widehat{s}_{i}\right) ^{2}=2%
\widehat{n}_{i}  \label{8}
\end{equation}%
which is independent of whether the system is doped by holes, and it shows
that the spin and charge degrees of electrons are intimately intertwined on
each lattice site, which is very important in the large $U$ limit for the
Hubbard model. For a Mott insulator described by the Hubbard model, we have
the relation, $\left( \widehat{n}_{i}\right) ^{2}=\left( \widehat{s}%
_{i}\right) ^{2}=1$; while for a doped Mott insulator with small hole doping
concentration $\delta $, we can take the approximation, called soft cut-off
approximation (SCA), in which we use the parameters $n^{2}=<\left( \widehat{n%
}_{i}\right) ^{2}>$ and $s^{2}=<\left( \widehat{s}_{i}\right) ^{2}>$ to
replace the charge and spin operators $\left( \widehat{n}_{i}\right) ^{2}$
and $\left( \widehat{s}_{i}\right) ^{2}$, respectively, in the Eqs.(\ref{3a1}%
-\ref{3a5}). These two parameters $n^{2}$ and $s^{2}$ have the relation with
the hole doping concentration $\delta $,%
\begin{equation}
n^{2}+s^{2}=2\left( 1-\delta \right)  \label{9}
\end{equation}%
which is rigorous due to the Eq.(\ref{8}). On the other hand, for enough of
large $U$ where the double occupied states are completely depleted, we can
take the relation, $n^{2}=s^{2}=1-\delta $.

Under the SCA, the equations of motion in Eqs.(\ref{3a3}-\ref{3a5}) can be
significantly simplified, but there are still some other high order ($L=3$)
multiple-point correlation functions. Fortunately, under the SCA, in these
equations there may appear the terms including the one-particle Green's
function. As a zeroth order approximation, we neglect all these high order ($%
L=3$) correlation functions appearing in the Eqs.(\ref{3a3}-\ref{3a5}), and
we only remain ones that are the lower order ($L\leq 2$) correlation
functions. Under this approximation, the set of the equations of motion of
the one-particle Green's function and the related correlation functions Eqs.(%
\ref{3a}-\ref{3a2}) is closed. In fact, this approximation is similar to a
"self-consistent field theory" that the multiple-point correlation functions 
$F_{iljq}^{(nn)}(\omega )$, $F_{iljq}^{(sn)}(\omega )$, $F_{iljq}^{(ss)}(%
\omega )$, $F_{ijlq}^{(P^{\left( -\right) })}(t)$ and $F_{ijlq}^{(Q^{\left(
-\right) })}(t)$ can be seen as some "external fields" that are the
functions of the one-particle Green's function and the correlation functions 
$F_{ilq}^{(n)}(t)$ as well as $F_{ilq}^{(s)}(t)$, then the set of equations
of motion in Eqs.(\ref{3a}-\ref{3a2}) can be self-consistently solved.

Obviously, the present approach is completely distinct from previous
perturbation expansion and/or cut-off approximations, and it has the
advantages that: (1) it can effectively treat the local spin and charge
fluctuations induced by the on-site repulsive Coulomb interaction with Eqs.(%
\ref{8},\ref{9}); (2) it effectively incorporate in the coupling between the
spin and charge fluctuations in Eqs.(\ref{3a1},\ref{3a2}); (3) it does not
need any "perturbation term" as a parameter to be expanded, like usual
perturbation expansion approach; (4) the basic equations of motion in Eqs.(%
\ref{3a}-\ref{3a2}) are rigorous, and the spin and charge fluctuations are
represented by the correlation functions $F_{ilq}^{(n)}(t)$ and $%
F_{ilq}^{(s)}(t)$, respectively, where they satisfy different equation of
motion. It is helpful in future to introducing different order parameters in
the low energy limit.

\section{Solution of the one-particle Green's function}

The related multiple-point correlation functions $F_{ijq}^{(n)}(t)$ and $%
F_{ijq}^{(s)}(t)$ appearing in the equation of the one-particle Green's
function are playing different roles, and they both have significant
contribution to the self-energy of electrons. Under the SCA, their
analytical expressions are given in the Appendix (\ref{a8a},\ref{a8b}). The
multiple-point correlation function $F_{ijq}^{(n)}(t)$ represents the
contribution of the charge fluctuation of electrons to the electronic
correlation effect. It has three significant contributions: (1) it modifies
the chemical potential by a factor, $-n^{2}J_{U}$; (2) it strongly
suppresses the spectral weight of the electrons by contributing a factor, $-<%
\widehat{n}_{i}>\delta _{iq}$ to the one-particle Green's function $%
G_{iq\sigma }(\omega )$, that makes the spectral weight of the electrons be
proportional to the doping concentration $\delta $; (3) it contributes a
term to the excitation spectrum proportional to $1-\delta $, that originates
from the coupling between the spin and charge fluctuations in the large $U$
limit. The multiple-point correlation function $F_{ijq}^{(s)}(t)$ represents
the contribution of the spin fluctuation of electrons to the electron
correlation effect. It significantly suppresses the itineracy of the
electrons, which makes the effective hopping amplitude of the electrons be
proportional to the hole doping concentration $\delta $. It means that the
strong spin fluctuation induced by the large on-site Coulomb interaction $U$
significantly suppresses the itineracy of the doping holes, and it tends to
localize the electrons.

With the help of the analytical expressions of the correlation functions $%
F_{ijq}^{(n)}(t)$ and $F_{ijq}^{(s)}(t)$, and after taking the Fourier
transformation of time, we can obtain the following equation of the
one-particle Green's function which is reliable in the low energy region ($%
\Delta <|\omega |\ll t_{0}$), 
\begin{equation}
\left[ \omega +\mu _{eff}\right] G_{iq\sigma }(\omega )=\left[ 1-<\widehat{n}%
_{i}>+\sigma <\widehat{s}_{i}>\right] \delta
_{iq}+\sum_{m}h_{im}^{eff}G_{mq\sigma }(\omega )  \label{10a}
\end{equation}%
where $\mu _{eff}=\mu -n^{2}J_{U}$, $J_{U}=\frac{4t_{0}^{2}}{U}$, and $%
h_{ij}^{eff}=\left( \delta +\frac{n^{2}J_{U}}{2U}\right) \widehat{h}_{ij}$.
If we take an average over the spin degrees of the one-particle Green's
function, $\overline{G}_{iq}(\omega )=\frac{1}{2}\left[ G_{iq\uparrow
}(\omega )+G_{iq\downarrow }(\omega )\right] $, the factor $\sigma <\widehat{%
s}_{i}>\delta _{iq}$ appearing in the Eq.(\ref{10a}) has no contribution to
the spectral weight of electrons. Thus, in the strong on-site repulsive
Coulomb interaction limit, $t_{0}\ll U$, the effective spectral weight of
electrons is that, $1-<\widehat{n}_{i}>=\delta $. Moreover, the equation of
motion of the one-particle Green's function in Eqs.(\ref{10a}) is valid for
a general lattice Hubbard model, for example, a chain, a square lattice, 
\textit{et al}.. On the other hand, as considering the next nearest neighbor
hoping of the electrons, $t_{ij}^{\prime }$, we should add a term $\delta
t_{im}^{\prime }G_{mq\sigma }(\omega )$ in the right side of Eq.(\ref{10a}),
where the parameter $t_{ij}^{\prime }$ is renormalized to $\delta
t_{ij}^{\prime }$ due the strong on-site repulsive Coulomb interaction.

After taking the Fourier transformation, we have the following analytical
expression of the one-particle Green's function in a square lattice%
\begin{equation}
\overline{G}_{k}(\omega )=\frac{\delta }{\omega +\mu _{eff}-\varepsilon
_{k}^{eff}}  \label{11}
\end{equation}%
where $\varepsilon _{k}^{eff}=\left( \delta +\frac{n^{2}J_{U}}{2U}\right)
\varepsilon _{k}^{0}$, and $\varepsilon _{k}^{0}=-2t_{0}\left[ \cos \left(
ak_{x}\right) +\cos \left( ak_{y}\right) \right] $, where $a$ is the lattice
constant. This dispersion $\varepsilon _{k}^{eff}$ is very similar to that
of the slave boson mean field theory with the complete condensed of holons
in the t-J model, and the term with the coefficient $\frac{n^{2}J_{U}}{2U}$
in the excitation spectrum may be corresponding to the excitation spectrum
of spinons\cite{13b}. This analytical expression of the one-particle Green's
function is reliable in the low energy region, $|\omega |\ll t_{0}$, as
calculating the contribution of the high order multiple-point correlation
functions we have used the conditions, $|\omega |\ll t_{0}\ll U$, $\mu
\simeq U/2$, and neglected the static quantities appearing in the equations
of motion of high order multiple-point correlation functions. That is, the
expression of the one-particle Green's function in the high energy region,
such as $|\omega |\gg t_{0}$, is different from the above one.

According to the Eq.(\ref{11}), the spectral function of electrons can be
written as that in the lower Hubbard band,%
\begin{equation}
A_{k}^{L}\left( \omega \right) =2\pi \delta \times \delta \left( \omega +\mu
_{eff}-\varepsilon _{k}^{eff}\right)  \label{12}
\end{equation}%
and the most of spectral weight is transferred to the upper Hubbard band in
the underdoping region. On the other hand, according to the related
multiple-point correlation functions $F_{ijq}^{(n)}(t)$ and $%
F_{ijq}^{(s)}(t) $, we can calculate the double occupation function of
electrons, $<\widehat{n}_{\uparrow i}\widehat{n}_{\downarrow i}>=-\frac{i}{2}%
\left[ F_{iii}^{(n)}(t)-\sigma F_{iii}^{(s)}(t)\right] _{t\rightarrow
-0^{+}} $, which is that in the lower Hubbard band,%
\begin{equation}
<\widehat{n}_{\uparrow i}\widehat{n}_{\downarrow i}>\propto \frac{\delta
\left( 1-\delta \right) t_{0}}{U}+O\left( \frac{1-\delta }{U^{2}}\right)
\label{13}
\end{equation}%
Obviously, this quantity goes to zero in the both regions around $\delta =0$
and $\delta =1$, respectively. The former is the result of the strong
on-site repulsive Coulomb interaction, in which the double occupied state on
each site is highly prohibited in the large $U$ limit, and the latter is
trivial due to no electron in the sites. Note that it is valid in the $%
U>>t_{0}$ limit due to we used the solutions of the correlation functions $%
F_{ijq}^{(n)}(t)$ and $F_{ijq}^{(s)}(t)$ in this limit.

At undoping $\delta =0$, where the system is at the half-filling of
electrons, the spectral weight of the electrons goes to zero, and the system
is a Mott insulator. Moreover, the excitation energy spectrum of electrons
is composed of two parts: one is proportional to the hole doping
concentration $\delta $, which goes to zero in the underdoping limit, and
another one is proportional to $1-\delta $, that take the maximum value at
undoping. \textit{Now it becomes more clear that the significantly
suppressed of the spectral weight (proportional to }$\delta $\textit{) of
electrons is mainly produced by the charge fluctuation described by the
correlation function }$F_{ijq}^{(n)}(t)$\textit{; while the localization
effect of electrons showing in the low energy excitation spectrum (the part
of it proportional to }$\delta $\textit{) is mainly produced by the spin
fluctuation represented by the correlation function }$F_{ijq}^{(s)}(t)$%
\textit{, and another part of the excitation spectrum proportional to }$%
1-\delta $\textit{\ is coming from the coupling between the charge and spin
fluctuations induced by the strong on-site repulsive Coulomb interaction. }%
Physically, it means that the charge and spin freedoms of electrons are
separated due to the strong on-site repulsive Coulomb interaction. This
phenomenon takes place not only in one dimension but also in two and three
dimensional lattices. The Mott insulator is a special case in which the
charge degrees of electrons is completely suppressed, and the low energy
excitation spectrum coming from the charge degrees of electrons is gapful.

The separation of the charge and spin degrees of freedom of the electrons
may be a common phenomenon in doped Mott insulators in which there exists
the strong correlation effect among electrons, and its origin is the strong
on-site repulsive Coulomb interaction of electrons. This is different from
that one for an one-dimensional weakly interacting electron gas where two
separated Fermi levels of the system play the key role in the charge-spin
separation of electrons, and even for very weak repulsive interaction, the
one-particle Green's function would show a power-law asymptotic behavior.
The present calculations are completely consistent with previous conjecture
that there takes place the charge and spin separation in strongly correlated
electronic systems which can be represented by the resonant valence bond
(RVB) states and/or slave-fermion/boson representations. In the doped Mott
insulators, the parameter of the hole doping concentration $\delta $ plays a
central role in understanding of the physical quantities observed in
experiments\cite{14} that are related to the one-particle Green's function,
such as the Drude weight, the Hall coefficient and the spectral function of
electrons, \textit{et al}.. For example, the present calculations show that
the spectral function of electrons around the Fermi surface is proportional
to $A_{k}^{L}\left( \omega \right) $, and the Drude weight\cite{15} is
proportional to $\delta $.

\section{Conclusion and discussion}

With the hierarchical Green's function approach, we have studied a doped
Mott insulator described by the Hubbard model by analytically solving the
equations of motion of the one-particle Green's function and related
multiple-point correlation functions, and found that the separation of the
spin and charge degrees of freedom of the electrons is an intrinsic
character of the doped Mott insulator. For enough of large on-site repulsive
Coulomb interaction, we have shown that the spectral weight of the
one-particle Green's function is proportional to the hole doping
concentration that is mainly produced by the charge fluctuation of the
electrons, while the excitation spectrum of the electrons is composed of two
parts: one is mainly contributed by the spin fluctuation of the electrons,
and its coefficient is proportional to the hole doping concentration that is
zero at the half-filling of the electrons; another one is coming from the
coupling between the charge and spin fluctuations of the electrons, which
can be seen as the spinon excitations, and its coefficient would decrease as
the hole doping concentration increasing. However, all these intrinsic low
energy properties of the system originate from the strong on-site repulsive
Coulomb interaction of electrons, that induces the strongly correlation
effect among electrons, and it may produce a variety of low energy
intertwined orders in different energy scales, such as spin density order,
charge density order, pre-pairing of electrons order and others.

The present results are consistent with the spectroscopy observations of the
cuprate superconductors in normal states\cite{8ad1,8ad2,8ad3}, and the
numerical calculations\cite{15a,15b} based on the Anderson's RVB theory with
a Gutzwiller projected BCS wave function. Moreover, this exotic excitation
spectrum of the electrons shows that only the doped holes take part in the
low energy/temperature transport behavior, and the electrons are nearly
localized in the underdoped regime, in which the spin fluctuation of the
electrons would play the important role in explaining the novel behavior of
the normal states in the underdoped and optimal doped regimes of the cuprate
superconductors.

The separation of the spin and charge degrees of freedom of the electrons in
a doped Mott insulator is derived from the competition between the tendency
to maintain local antiferromagnetic correlations originated from the strong
on-site repulsive Coulomb interaction and the doped hole itineracy, and this
strong correlation effect of the electrons would induce some intertwined
orders in the low energy region. To deeply understanding of the low energy
property of a doped Mott insulator, we need to introduce these possible
orders, and to judge which one of them can survive and be robust in the
strong on-site repulsive Coulomb interaction. The recent calculations of the
spin susceptibility for a spin $1/2$ Heisenberg model on a square lattice
show that the low-lying excitations of the spins can be divided as two
parts: one is the spin wave excitations that residing in the lowest boundary
of the low-lying excitations around the momentum $\mathbf{k}=(\pi ,\pi )$;
another one is the nearly deconfined spinon excitations that mainly residing
in the high energy region around the momentum $\mathbf{k}=(\pi ,\pi )$ and
other momentum regions, such as, $\mathbf{k}=(\pi ,0)$ and $\mathbf{k}=(%
\frac{\pi }{2},\frac{\pi }{2})$, \textit{et al}., where the spin wave
excitations become very weak. This phenomenon is observed in a recent
neutron scattering experiment\cite{16}, and numerical calculations\cite{17}.
The influence of these two distinct low-lying excitations of the spins on
the intertwined orders observed in the normal state of the cuprate
superconductors deserves to be further carefully studied in the future.

\section{Acknowledgments}

This work is supported by the National Natural Science Foundation of China
under Grant No. 11074301, and the National Basic Research Program of China
under Grant No. 2012CB921704.

\section{Appendix}

The Hamiltonian of the Hubbard model is that,%
\begin{equation}
\widehat{H}_{H}=-t_{0}\sum_{ij\sigma }\widehat{\gamma }_{ij}\left( \widehat{c%
}_{\sigma i}^{\dagger }\widehat{c}_{\sigma j}+\widehat{c}_{\sigma
j}^{\dagger }\widehat{c}_{\sigma i}\right) +U\sum_{i}\widehat{n}_{\uparrow i}%
\widehat{n}_{\downarrow i}-\mu \sum_{i\sigma }\widehat{n}_{\sigma i}
\label{1}
\end{equation}%
where $\widehat{c}_{\sigma i}^{\dagger }$/$\widehat{c}_{\sigma i}$
creates/annihilates an electron with spin $\sigma =\uparrow $, $\downarrow $
on site $\mathbf{x}_{i}$, $\widehat{n}_{\sigma i}=\widehat{c}_{\sigma
i}^{\dagger }\widehat{c}_{\sigma i}$ is the number operator, $\mu $ is the
chemical potential, $t_{0}$ is the hopping amplitude, and $U$ is the on-site
repulsive Coulomb interaction strength. The hopping factor $\widehat{\gamma }%
_{ij}$ is defined as that,%
\begin{equation*}
\widehat{\gamma }_{ij}=\left\{ 
\begin{array}{cc}
1, & j=i+1 \\ 
0, & j\neq i+1%
\end{array}%
\right.
\end{equation*}%
which denotes the summation over the sites $\mathbf{x}_{i},\mathbf{x}_{j}$
only in the nearest neighbor. Here we take the on-site repulsive Coulomb
interaction $U$ as one of the largest energy scale of the system, which is
of order the bandwidth $W$($=8t_{0}$), $t_{0}\ll U\sim W$.

\subsection{The basic commutation relations}

According to the definitions of the one-particle Green's function, $%
G_{ij}(t)=-i<\mathit{T}\widehat{c}_{\sigma i}(t)\widehat{c}_{\sigma
j}^{\dagger }(0)>$, in order to write out its equation of motion with the
hierarchic Green's function approach, we need the commutation relations of
the electron operators $\widehat{c}_{\sigma i}$ and other operators with the
Hamiltonian, here we write out some of them as that, 
\begin{equation}
\left[ \widehat{c}_{\sigma i},H\right] =\widehat{h}_{im}\widehat{c}_{\sigma
m}-\mu \widehat{c}_{\sigma i}+\frac{U}{2}\left[ \widehat{n}_{i}-\sigma 
\widehat{s}_{i}\right] \widehat{c}_{\sigma i}  \label{2a}
\end{equation}%
\begin{eqnarray}
\lbrack \widehat{n}_{i},\widehat{H}] &=&\widehat{h}_{im}\widehat{P}%
_{im}^{\left( -\right) }  \notag \\
\lbrack \widehat{s}_{i},\widehat{H}] &=&\widehat{h}_{im}\widehat{Q}%
_{im}^{\left( -\right) }  \label{2c}
\end{eqnarray}%
where $\widehat{h}_{il}=-t_{0}\left( \widehat{\gamma }_{il}+\widehat{\gamma }%
_{li}\right) $, $\widehat{n}_{i}=\widehat{n}_{\uparrow i}+\widehat{n}%
_{\downarrow i}$, $\widehat{s}_{i}=\widehat{n}_{\uparrow i}-\widehat{n}%
_{\downarrow i}$, $\widehat{P}_{ij}^{\left( \pm \right) }=\widehat{N}%
_{ij}\pm \widehat{N}_{ji}$, and $\widehat{Q}_{ij}^{\left( \pm \right) }=%
\widehat{S}_{ij}\pm \widehat{S}_{ji}$, where $\widehat{N}_{ij}=\widehat{c}%
_{\uparrow i}^{\dagger }\widehat{c}_{\uparrow j}+\widehat{c}_{\downarrow
i}^{\dagger }\widehat{c}_{\downarrow j}$, and $\widehat{S}_{ij}=\widehat{c}%
_{\uparrow i}^{\dagger }\widehat{c}_{\uparrow j}-\widehat{c}_{\downarrow
i}^{\dagger }\widehat{c}_{\downarrow j}$. These commutation relations are
the elementary ingredients as writing out the equations of motion of related
multiple-point correlation functions that directly or indirectly appearing
in the equation of motion of the one-particle Green's function. Here we
separate the charge and spin degrees of freedom of electrons by writing out, 
$\widehat{n}_{\sigma i}=\frac{1}{2}\left[ \widehat{n}_{i}-\sigma \widehat{s}%
_{i}\right] $, that is produced by the on-site Coulomb interaction in the
above commutation relations. In this way, we can separately define the
charge and spin fluctuations of electrons by different correlation functions.

\subsection{Definition of multiple-point correlation functions}

In order to tersely represent multiple-point correlation functions, we
introduce new composite multiple-point operators $\widehat{F}_{\{\alpha
_{1}...\alpha _{L}\}}^{(A_{1}\cdots A_{L})}$,

\begin{equation}
\widehat{F}_{\{\alpha _{1}...\alpha _{L}\}}^{(A_{1}\cdots A_{L})}=\underset{%
k=1}{\Pi }^{L}\widehat{A}_{\alpha _{k}}  \label{5}
\end{equation}%
where $\widehat{A}_{\alpha }=\left\{ \widehat{n}_{i},\widehat{s}_{i},%
\widehat{P}_{ij}^{(\pm )},\widehat{Q}_{ij}^{(\pm )}\right\} $, and $L$ is
the number of the operator $\widehat{A}_{\alpha }$ appearing in the
composite operators $\widehat{F}_{\{\alpha _{1}...\alpha
_{L}\}}^{(A_{1}\cdots A_{L})}$. With the help of these composite operators $%
\widehat{F}_{\{\alpha _{1}...\alpha _{L}\}}^{(A_{1}\cdots A_{L})}$, we
define the corresponding multiple-point correlation functions $F_{\{\alpha
_{1}...\alpha _{L}\}mq}^{(A_{1}\cdots A_{L})}(t_{1},t_{2})$, 
\begin{equation}
F_{\{\alpha _{1}...\alpha _{L}\}mq}^{(A_{1}\cdots A_{L})}(t)=-i<\mathit{T}%
\widehat{F}_{\{\alpha _{1}...\alpha _{L}\}}^{(A_{1}\cdots A_{L})}(t)\widehat{%
c}_{m\sigma }(t)\widehat{c}_{q\sigma }^{\dagger }(0)>  \label{6}
\end{equation}%
Some of these multiple-point correlation functions $F_{\{\alpha
_{1}...\alpha _{L}\}mq}^{(A_{1}\cdots A_{L})}(t)$ will enter into the series
of hierarchical equations of motion originated from the equation of motion
of the one-particle Green's function Eqs.(\ref{3a}), such as $%
F_{ijq}^{(n)}(t)$ and $F_{ijq}^{(s)}(t)$, \textit{et al}., and they will
construct a set of linear equations of motion with the one-particle Green's
function $G_{iq\sigma }(t)$. The physics meaning, for example, of the
operator $\widehat{F}_{\{\alpha _{1}...\alpha _{L}\}}^{(\widehat{A}%
_{1}\cdots \widehat{A}_{L})}\widehat{c}_{m\sigma }$ in the correlation
function $F_{\{\alpha _{1}...\alpha _{L}\}mq}^{(A_{1}\cdots A_{L})}(t)$ is
that an electron $\widehat{c}_{m\sigma }$ with spin $\sigma $ at site $%
\mathbf{x}_{m}$ attached other electrons represented by the operator $%
\widehat{F}_{\{\alpha _{1}...\alpha _{L}\}}^{(\widehat{A}_{1}\cdots \widehat{%
A}_{L})}$ around the site $\mathbf{x}_{m}$, where the parameter $L$ denotes
the number of electrons residing in a length (sites) scale around this site $%
\mathbf{x}_{m}$ that the electrons in this scale all are involved in the
time evolution process of the electron $\widehat{c}_{m\sigma }$. Thus the
correlation function $F_{\{\alpha _{1}...\alpha _{L}\}mq}^{(A_{1}\cdots
A_{L})}(t)$ in fact represents the evolution process of an electron from the
initial state incorporated the influence of a definite distribution of other
electrons around it to final state.

In contrast with usual correlation functions defined in the momentum space,
the present multiple-point correlation functions can more effectively
describe the correlation effect of electrons derived from the on-site
Coulomb interaction, and the parameter $L$ appearing in the composite
operators $\widehat{F}_{\{\alpha _{1}...\alpha _{L}\}}^{(\widehat{A}%
_{1}\cdots \widehat{A}_{L})}$ can be used to classify the correlation
functions $F_{\{\alpha _{1}...\alpha _{L}\}mq}^{(A_{1}\cdots A_{L})}(t)$
into different levels, where the correlation functions $F_{\{\alpha
_{1}...\alpha _{L}\}mq}^{(A_{1}\cdots A_{L})}(t)$ in the same level $L$ can
constitute one or more subset of equations of motion.

\subsection{Solutions of the high order related multiple-point correlation
functions}

Based on the basic commutation relations in the Eqs.(\ref{2a},\ref{2c}), we
can write out equations of motion of the high order related multiple-point
correlation functions,%
\begin{eqnarray}
\left[ \omega +\mu \right] F_{ilq}^{(n)}(\omega ) &=&<\widehat{n}_{i}>\delta
_{lq}+\frac{U}{2}F_{illq}^{(nn)}(\omega )-\frac{\sigma U}{2}%
F_{illq}^{(ns)}(\omega )  \notag \\
&&+\sum_{m}\left[ \widehat{h}_{lm}F_{imq}^{(n)}(\omega )+\widehat{h}%
_{im}F_{imlq}^{(P^{\left( -\right) })}(\omega )\right]  \label{a1a}
\end{eqnarray}%
\begin{eqnarray}
\left[ \omega +\mu \right] F_{ilq}^{(s)}(\omega ) &=&<\widehat{s}_{i}>\delta
_{lq}+\frac{U}{2}F_{illq}^{(sn)}(\omega )-\frac{\sigma U}{2}%
F_{illq}^{(ss)}(\omega )  \notag \\
&&+\sum_{m}\left[ \widehat{h}_{lm}F_{imq}^{(s)}(\omega )+\widehat{h}%
_{im}F_{imlq}^{(Q^{\left( -\right) })}(\omega )\right]  \label{a1b}
\end{eqnarray}%
\begin{eqnarray}
\left[ \omega +\mu \right] F_{iljq}^{(nn)}(\omega ) &=&<\widehat{n}_{i}%
\widehat{n}_{l}>\delta _{jq}+\frac{U}{2}F_{iljjq}^{(nnn)}(\omega )-\frac{%
\sigma U}{2}F_{iljjq}^{(nns)}(\omega )  \notag \\
&&+\sum_{m}\left[ \widehat{h}_{jm}F_{ilmq}^{(nn)}(\omega )+\widehat{h}%
_{im}F_{imljq}^{(P^{\left( -\right) }n)}(\omega )+\widehat{h}%
_{lm}F_{ilmjq}^{(nP^{\left( -\right) })}(\omega )\right]  \label{a2a}
\end{eqnarray}%
\begin{eqnarray}
\left[ \omega +\mu \right] F_{iljq}^{(sn)}(\omega ) &=&<\widehat{s}_{i}%
\widehat{n}_{l}>\delta _{jq}+\frac{U}{2}F_{iljjq}^{(snn)}(\omega )-\frac{%
\sigma U}{2}F_{iljjq}^{(sns)}(\omega )  \notag \\
&&+\sum_{m}\left[ \widehat{h}_{jm}F_{ilmq}^{(sn)}(\omega )+\widehat{h}%
_{im}F_{imljq}^{(Q^{\left( -\right) }n)}(\omega )+\widehat{h}%
_{lm}F_{ilmjq}^{(sP^{\left( -\right) })}(\omega )\right]  \label{a2b}
\end{eqnarray}%
\begin{eqnarray}
\left[ \omega +\mu \right] F_{iljq}^{(ns)}(\omega ) &=&<\widehat{n}_{i}%
\widehat{s}_{l}>\delta _{jq}+\frac{U}{2}F_{iljjq}^{(nsn)}(\omega )-\frac{%
\sigma U}{2}F_{iljjq}^{(nss)}(\omega )  \notag \\
&&+\sum_{m}\left[ \widehat{h}_{jm}F_{ilmq}^{(ns)}(\omega )+\widehat{h}%
_{im}F_{imljq}^{(P^{\left( -\right) }s)}(\omega )+\widehat{h}%
_{lm}F_{ilmjq}^{(nQ^{\left( -\right) })}(\omega )\right]  \label{a2c}
\end{eqnarray}%
\begin{eqnarray}
\left[ \omega +\mu \right] F_{iljq}^{(ss)}(\omega ) &=&<\widehat{s}_{i}%
\widehat{s}_{l}>\delta _{jq}+\frac{U}{2}F_{iljjq}^{(ssn)}(\omega )-\frac{%
\sigma U}{2}F_{iljjq}^{(sss)}(\omega )  \notag \\
&&+\sum_{m}\left[ \widehat{h}_{jm}F_{ilmq}^{(ss)}(\omega )+\widehat{h}%
_{im}F_{imljq}^{(Q^{\left( -\right) }s)}(\omega )+\widehat{h}%
_{lm}F_{ilmjq}^{(sQ^{\left( -\right) })}(\omega )\right]  \label{a2d}
\end{eqnarray}%
where we have written out the equations of motion of the related
multiple-point correlation functions $F_{iljq}^{(sn)}(\omega )$ and $%
F_{iljq}^{(ns)}(\omega )$, respectively, that would be different under the
SCA approximation due to the different order of the operators $\widehat{s}%
_{i}$ and $\widehat{n}_{i}$ appearing in these correlation functions. With
the same procedures, we can also write out the equations of motion of the
correlation functions $F_{ijlq}^{(P^{\left( -\right) })}(\omega )$ and $%
F_{ijlq}^{(Q^{\left( -\right) })}(\omega )$ where their contributions are
mainly to modify the chemical potential of electrons. Here we do not further
to write out the equations of motion of the multiple-point correlation
functions belonging to the $L=3$ level, such as $F_{iljjq}^{(nnn)}(\omega )$%
, $F_{iljjq}^{(nns)}(\omega )$, and $F_{iljjq}^{(sss)}(\omega )$, et al.,
and we take a cut-off approximation, that is under the SCA approximation we
simply discard those multiple-point correlation functions belonging to the $%
L=3$ level.

After taking the SCA approximation, and only keeping the lowest order terms,
we can write out the equations of motion of the correlation functions $%
F_{iiq}^{(n)}(\omega )$ and $F_{iiq}^{(s)}(\omega )$ as that (taking $\mu
=U/2$),%
\begin{eqnarray}
\left[ \Omega -\eta \left( \omega \right) \right] F_{iiq}^{(n)}(\omega ) &=&<%
\widehat{n}_{i}>\delta _{iq}-\frac{Us^{2}}{2}G_{iq\sigma }(\omega )  \notag
\\
&&+\sum_{m}\widehat{h}_{im}F_{imq}^{(n)}(\omega )-\frac{\sigma U}{2}%
F_{iiq}^{(s)}(\omega )  \label{a4a}
\end{eqnarray}%
\begin{equation}
\left[ \omega -\eta \left( \omega \right) \right] F_{iiq}^{(s)}(\omega )=<%
\widehat{s}_{i}>\delta _{iq}-\frac{\sigma Us^{2}}{2}G_{iq\sigma }(\omega
)+\sum_{m}\widehat{h}_{im}F_{imq}^{(s)}(\omega )  \label{a4b}
\end{equation}%
\begin{equation}
\left[ \omega +\frac{U}{2}\right] F_{ijq}^{(n)}(\omega )=<\widehat{n}%
_{i}>\delta _{jq}+\frac{U}{2}F_{ijjq}^{(nn)}(\omega )-\frac{\sigma U}{2}%
F_{ijjq}^{(ns)}(\omega )  \label{a5a}
\end{equation}%
\begin{equation}
\left[ \omega +\frac{U}{2}\right] F_{ijq}^{(s)}(\omega )=<\widehat{s}%
_{i}>\delta _{jq}+\frac{U}{2}F_{ijjq}^{(sn)}(\omega )-\frac{\sigma U}{2}%
F_{ijjq}^{(ss)}(\omega )  \label{a5b}
\end{equation}%
where $\eta \left( \omega \right) =\frac{2}{\omega +\frac{U}{2}}%
\sum_{m}\left( \widehat{h}_{im}\right) ^{2}$, it is contributed by the
correlation functions $F_{ijlq}^{(P^{\left( -\right) })}(\omega )$ or $%
F_{ijlq}^{(Q^{\left( -\right) })}(\omega )$. The other equations of motion
of the high order related multiple-point correlation functions can be
reduced as that after neglecting high order terms,%
\begin{eqnarray}
\left[ \omega -\frac{U}{2}\right] F_{ijjq}^{(nn)}(\omega ) &=&<\widehat{n}%
_{i}\widehat{n}_{j}>\delta _{jq}+\widehat{h}_{ji}F_{ijiq}^{(nn)}(\omega )-%
\frac{s^{2}U}{2}F_{ijq}^{(n)}(\omega )  \notag \\
&&-\frac{1}{2}\widehat{h}_{ij}\left[ F_{jiq}^{(n)}(\omega )-n^{2}G_{iq\sigma
}(\omega )\right]  \label{a6a}
\end{eqnarray}%
\begin{eqnarray}
\omega F_{ijjq}^{(ns)}(\omega ) &=&<\widehat{n}_{i}\widehat{s}_{j}>\delta
_{jq}+\frac{\sigma }{2}\widehat{h}_{ij}\left[ F_{jiq}^{(n)}(\omega
)-n^{2}G_{iq\sigma }(\omega )\right]  \notag \\
&&+\widehat{h}_{ji}F_{ijiq}^{(ns)}(\omega )-\frac{\sigma s^{2}U}{2}%
F_{ijq}^{(n)}(\omega )  \label{a6b}
\end{eqnarray}%
\begin{eqnarray}
\left[ \omega -\frac{U}{2}\right] F_{ijiq}^{(ns)}(\omega ) &=&<\widehat{n}%
_{i}\widehat{s}_{j}>\delta _{iq}+\widehat{h}_{ij}F_{ijjq}^{(ns)}(\omega )-%
\frac{s^{2}U}{2}F_{jiq}^{(s)}(\omega )  \notag \\
&&+\frac{\sigma }{2}\widehat{h}_{ji}\left[ n^{2}G_{jq\sigma }(\omega
)+\sigma F_{ijq}^{(s)}(\omega )\right]  \label{a6c}
\end{eqnarray}%
\begin{eqnarray}
\left[ \omega -\frac{U}{2}\right] F_{ijjq}^{(sn)}(\omega ) &=&<\widehat{s}%
_{i}\widehat{n}_{j}>\delta _{jq}+\widehat{h}_{ji}F_{ijiq}^{(sn)}(\omega )-%
\frac{s^{2}U}{2}F_{ijq}^{(s)}(\omega )  \notag \\
&&-\frac{\sigma }{2}\widehat{h}_{ij}\left[ F_{jiq}^{(n)}(\omega
)-n^{2}G_{iq\sigma }(\omega )\right]  \label{a6d}
\end{eqnarray}%
\begin{eqnarray}
\omega F_{ijiq}^{(sn)}(\omega ) &=&<\widehat{s}_{i}\widehat{n}_{j}>\delta
_{iq}+\widehat{h}_{ij}F_{ijjq}^{(sn)}(\omega )-\frac{\sigma s^{2}U}{2}%
F_{jiq}^{(n)}(\omega )  \notag \\
&&+\frac{1}{2}\widehat{h}_{ji}\left[ \sigma s^{2}G_{jq\sigma }(\omega
)+F_{ijq}^{(s)}(\omega )\right]  \label{a6e}
\end{eqnarray}%
\begin{eqnarray}
\omega F_{ijjq}^{(ss)}(\omega ) &=&<\widehat{s}_{i}\widehat{s}_{j}>\delta
_{jq}+\frac{1}{2}\widehat{h}_{ij}\left[ F_{jiq}^{(n)}(\omega
)-n^{2}G_{iq\sigma }(\omega )\right]  \notag \\
&&+\widehat{h}_{ji}F_{ijiq}^{(ss)}(\omega )-\frac{\sigma s^{2}U}{2}%
F_{ijq}^{(s)}(\omega )  \label{a6f}
\end{eqnarray}%
where under the present approximation the difference between $%
F_{ijjq}^{(ns)}(\omega )$ and $F_{ijiq}^{(sn)}(\omega )$, as well as between 
$F_{ijiq}^{(ns)}(\omega )$ and $F_{ijjq}^{(sn)}(\omega )$ originates from
the different order of the operators $\widehat{s}_{i}$ and $\widehat{n}_{i}$
appearing in these correlation functions. In the above calculations, we
approximately take the chemical potential, $\mu =U/2$, that is rigorous at
undoping, and reasonable in the underdoped region.

These equations in Eqs.(\ref{a6a}-\ref{a6f}) are some simple sets of
equations, and they can be analytically solved as that,%
\begin{eqnarray}
F_{ijjq}^{(nn)}(\omega ) &=&\frac{\Omega }{D_{nn}\left( \omega \right) }%
\left\{ <\widehat{n}_{i}\widehat{n}_{j}>\delta _{jq}-\frac{s^{2}U}{2}%
F_{ijq}^{(n)}(\omega )-\frac{1}{2}\widehat{h}_{ij}\left[ F_{jiq}^{(n)}(%
\omega )-n^{2}G_{iq\sigma }(\omega )\right] \right\}  \notag \\
&&+\frac{\widehat{h}_{ij}}{D_{nn}\left( \omega \right) }\left\{ <\widehat{n}%
_{i}\widehat{n}_{j}>\delta _{iq}-\frac{s^{2}U}{2}F_{jiq}^{(n)}(\omega )-%
\frac{1}{2}\widehat{h}_{ji}\left[ F_{ijq}^{(n)}(\omega )-n^{2}G_{jq\sigma
}(\omega )\right] \right\}  \label{a7a}
\end{eqnarray}%
\begin{eqnarray}
F_{ijjq}^{(ns)}(\omega ) &=&\frac{\Omega }{D_{ns}\left( \omega \right) }%
\left\{ <\widehat{n}_{i}\widehat{s}_{j}>\delta _{jq}-\frac{\sigma s^{2}U}{2}%
F_{ijq}^{(n)}(\omega )+\frac{\sigma }{2}\widehat{h}_{ij}\left[
F_{jiq}^{(n)}(\omega )-n^{2}G_{iq\sigma }(\omega )\right] \right\}  \notag \\
&&+\frac{\widehat{h}_{ij}}{D_{ns}\left( \omega \right) }\left\{ <\widehat{n}%
_{i}\widehat{s}_{j}>\delta _{iq}-\frac{s^{2}U}{2}F_{jiq}^{(s)}(\omega
)+\sigma \widehat{h}_{li}\left[ \frac{n^{2}}{2}G_{lq\sigma }(\omega )+\frac{%
\sigma }{2}F_{ilq}^{(s)}(\omega )\right] \right\}  \label{a7b}
\end{eqnarray}%
\begin{eqnarray}
F_{ijjq}^{(sn)}(\omega ) &=&\frac{\omega }{D_{ns}\left( \omega \right) }%
\left\{ <\widehat{n}_{i}\widehat{s}_{j}>\delta _{jq}-\frac{\sigma s^{2}U}{2}%
F_{ijq}^{(n)}(\omega )+\frac{\sigma }{2}\widehat{h}_{ij}\left[
F_{jiq}^{(n)}(\omega )-n^{2}G_{iq\sigma }(\omega )\right] \right\}  \notag \\
&&+\frac{\widehat{h}_{ij}}{D_{ns}\left( \omega \right) }\left\{ <\widehat{n}%
_{i}\widehat{s}_{j}>\delta _{iq}-\frac{s^{2}U}{2}F_{jiq}^{(s)}(\omega )+%
\frac{1}{2}\widehat{h}_{ji}\left[ \sigma s^{2}G_{jq\sigma }(\omega
)+F_{ijq}^{(s)}(\omega )\right] \right\}  \label{a7c}
\end{eqnarray}%
\begin{eqnarray}
F_{ijjq}^{(ss)}(\omega ) &=&\frac{\omega }{D_{ss}\left( \omega \right) }%
\left\{ <\widehat{s}_{i}\widehat{s}_{j}>\delta _{jq}-\frac{\sigma s^{2}U}{2}%
F_{ijq}^{(s)}(\omega )+\frac{1}{2}\widehat{h}_{ij}\left[ \sigma
F_{jiq}^{(n)}(\omega )-n^{2}G_{iq\sigma }(\omega )\right] \right\}  \notag \\
&&+\frac{\widehat{h}_{ij}}{D_{ss}\left( \omega \right) }\left\{ <\widehat{s}%
_{i}\widehat{s}_{j}>\delta _{iq}-\frac{\sigma s^{2}U}{2}F_{jiq}^{(s)}(\omega
)-\frac{1}{2}\widehat{h}_{ji}\left[ F_{ijq}^{(s)}(\omega )+n^{2}G_{jq\sigma
}(\omega )\right] \right\}  \label{a7d}
\end{eqnarray}%
where $\Omega =\omega -\frac{U}{2}$, $D_{nn}\left( \omega \right) =\Omega
^{2}-t_{0}^{2}$, $D_{ns}\left( \omega \right) =\omega \Omega -t_{0}^{2}$,
and $D_{ss}\left( \omega \right) =\omega ^{2}-t_{0}^{2}$. Obviously, under
the present approximation, these high order correlation functions are the
functional of the one-particle Green's function and the correlation
functions $F_{ijq}^{(n)}(\omega )$ and $F_{ijq}^{(s)}(\omega )$ that
represent the charge and spin fluctuations, respectively. While they can be
seen as some "effective external fields" that appear in the equations of
motion of the correlation functions $F_{ijq}^{(n)}(\omega )$ and $%
F_{ijq}^{(s)}(\omega )$ in the Eqs.(\ref{a5a},\ref{a5b}).

Substituting these solutions in Eqs.(\ref{a7a}-\ref{a7d}) into the Eqs.(\ref%
{a5a},\ref{a5b}), we have the following expressions of the correlation
functions $F_{ijq}^{(n)}(\omega )$ and $F_{ijq}^{(s)}(\omega )$, 
\begin{eqnarray}
F_{ijq}^{(n)}(\omega ) &=&\frac{<\widehat{n}_{i}>\delta _{jq}}{\Gamma
_{n}\left( \omega \right) }+\frac{U}{2\Gamma _{n}\left( \omega \right) }%
\frac{<\widehat{n}_{i}\widehat{n}_{j}>\left[ \Omega \delta _{jq}+\widehat{h}%
_{ij}\delta _{iq}\right] }{D_{nn}\left( \omega \right) }  \notag \\
&&+\frac{n^{2}\Omega U}{4\Gamma _{n}\left( \omega \right) }\left( \frac{1}{%
D_{nn}\left( \omega \right) }+\frac{1}{D_{ns}\left( \omega \right) }\right) 
\widehat{h}_{ij}G_{iq\sigma }(\omega )  \label{a8a} \\
&&+\frac{n^{2}U}{4\Gamma _{n}\left( \omega \right) }\left( \frac{1}{%
D_{nn}\left( \omega \right) }-\frac{1}{D_{ns}\left( \omega \right) }\right)
\left( \widehat{h}_{ij}\right) ^{2}G_{jq\sigma }(\omega )  \notag
\end{eqnarray}%
\begin{eqnarray}
F_{ijq}^{(s)}(\omega ) &=&\frac{<\widehat{s}_{i}>\delta _{jq}}{\Gamma
_{s}\left( \omega \right) }-\frac{U}{2\Gamma _{s}\left( \omega \right) }%
\frac{\sigma <\widehat{s}_{i}\widehat{s}_{j}>\left[ \omega \delta _{jq}+%
\widehat{h}_{ij}\delta _{iq}\right] }{D_{ss}\left( \omega \right) }  \notag
\\
&&+\frac{\sigma n^{2}\omega U}{4\Gamma _{s}\left( \omega \right) }\left( 
\frac{1}{D_{ss}\left( \omega \right) }-\frac{1}{D_{ns}\left( \omega \right) }%
\right) \widehat{h}_{ij}G_{iq\sigma }(\omega )  \label{a8b} \\
&&+\frac{\sigma U}{4\Gamma _{s}\left( \omega \right) }\left( \frac{n^{2}}{%
D_{ss}\left( \omega \right) }+\frac{s^{2}}{D_{ns}\left( \omega \right) }%
\right) \left( \widehat{h}_{ji}\right) ^{2}G_{jq\sigma }(\omega )  \notag
\end{eqnarray}%
where $\Gamma _{n}\left( \omega \right) =\omega +\frac{U}{2}\left( 1-\frac{%
s^{2}\Omega U}{2D_{ns}\left( \omega \right) }\right) $ and $\Gamma
_{s}\left( \omega \right) =\omega +\frac{U}{2}\left( 1-\frac{s^{2}\omega U}{%
2D_{ss}\left( \omega \right) }\right) $. With the above expressions and the
Eqs.(\ref{a4a},\ref{a4b}), we finally obtain the following analytical
expression of the correlation function $F_{iiq}^{(n)}(\omega )-\sigma
F_{iiq}^{(s)}(\omega )$ by the one-particle Green's function,%
\begin{eqnarray}
F_{iiq}^{(n)}(\omega )-\sigma F_{iiq}^{(s)}(\omega ) &=&\frac{<\widehat{n}%
_{i}>\delta _{iq}+P_{iq}^{\left( n\right) }\left( \omega \right) }{\Omega
-\eta \left( \omega \right) }-\frac{\sigma \left[ <\widehat{s}_{i}>\delta
_{iq}+P_{iq\sigma }^{\left( s\right) }\left( \omega \right) \right] }{\Omega
-\eta \left( \omega \right) }  \notag \\
&&+\alpha \left( \omega \right) G_{iq\sigma }(\omega )+\sum_{m}\zeta
_{im}\left( \omega \right) G_{mq\sigma }(\omega )  \label{a9}
\end{eqnarray}%
where the coefficients $\alpha \left( \omega \right) $, $\zeta _{im}\left(
\omega \right) $, $P_{iq}^{\left( n\right) }\left( \omega \right) $ and $%
P_{iq\sigma }^{\left( s\right) }\left( \omega \right) $ read that,%
\begin{eqnarray}
\alpha \left( \omega \right) &=&\frac{n^{2}\Omega U}{4\Gamma _{n}\left(
\omega \right) \left[ \Omega -\eta \left( \omega \right) \right] }\left( 
\frac{1}{D_{nn}\left( \omega \right) }+\frac{1}{D_{ns}\left( \omega \right) }%
\right) \sum_{m}\left( \widehat{h}_{im}\right) ^{2}  \notag \\
&&-\frac{n^{2}\omega U}{4\Gamma _{s}\left( \omega \right) \left[ \Omega
-\eta \left( \omega \right) \right] }\left( \frac{1}{D_{ss}\left( \omega
\right) }-\frac{1}{D_{ns}\left( \omega \right) }\right) \sum_{m}\left( 
\widehat{h}_{im}\right) ^{2}  \label{a10a}
\end{eqnarray}%
\begin{eqnarray}
\zeta _{im}\left( \omega \right) &=&\frac{n^{2}t_{0}^{2}U}{4\Gamma
_{n}\left( \omega \right) \left[ \Omega -\eta \left( \omega \right) \right] }%
\left( \frac{1}{D_{nn}\left( \omega \right) }-\frac{1}{D_{ns}\left( \omega
\right) }\right) \widehat{h}_{im}  \notag \\
&&-\frac{t_{0}^{2}U}{4\Gamma _{s}\left( \omega \right) \left[ \Omega -\eta
\left( \omega \right) \right] }\left( \frac{n^{2}}{D_{ss}\left( \omega
\right) }+\frac{s^{2}}{D_{ns}\left( \omega \right) }\right) \widehat{h}_{im}
\label{a10b}
\end{eqnarray}%
\begin{equation}
P_{iq}^{\left( n\right) }\left( \omega \right) =\sum_{m}\left( \frac{<%
\widehat{n}_{i}>\widehat{h}_{im}\delta _{mq}}{\Gamma _{n}\left( \omega
\right) }+\frac{U}{2\Gamma _{n}\left( \omega \right) }\frac{<\widehat{n}_{i}%
\widehat{n}_{m}>\widehat{h}_{im}\left[ \Omega \delta _{mq}+\widehat{h}%
_{im}\delta _{iq}\right] }{D_{nn}\left( \omega \right) }\right)  \label{a10c}
\end{equation}%
\begin{equation}
P_{iq\sigma }^{\left( s\right) }\left( \omega \right) =\sum_{m}\left( \frac{<%
\widehat{s}_{i}>\widehat{h}_{im}\delta _{mq}}{\Gamma _{s}\left( \omega
\right) }-\frac{U}{2\Gamma _{s}\left( \omega \right) }\frac{\sigma <\widehat{%
s}_{i}\widehat{s}_{m}>\widehat{h}_{im}\left[ \omega \delta _{mq}+\widehat{h}%
_{im}\delta _{iq}\right] }{D_{ss}\left( \omega \right) }\right)  \label{a10d}
\end{equation}%
In the low energy limit, $\omega \rightarrow 0$, the coefficients $\alpha
\left( \omega \right) $ and $\zeta _{im}\left( \omega \right) $ are reduced
as that,%
\begin{equation}
\zeta _{im}\left( \omega \right) \overset{\omega \rightarrow 0}{=}\left[ 
\frac{4n^{2}t_{0}^{2}}{U^{3}}-\frac{n^{2}+s^{2}}{U}\right] \widehat{h}_{im}
\label{a11a}
\end{equation}%
\begin{equation}
\alpha \left( \omega \right) \overset{\omega \rightarrow 0}{=}\frac{n^{2}}{8}%
\left( \frac{W_{2D}}{U}\right) ^{2}  \label{a11b}
\end{equation}%
where $W_{2D}=8t_{0}$. With the Eq.(\ref{a9}), and neglecting the functions $%
P_{iq}^{\left( n\right) }\left( \omega \right) $ and $P_{iq\sigma }^{\left(
s\right) }\left( \omega \right) $, we can obtain the equation of motion of
the one-particle Green's function in Eq.(\ref{10a}) under the large $U$
limit, $|\omega |\ll t_{0}\ll U\sim W$.

\section{\protect\bigskip}

\end{document}